\documentclass[aps,prb,twocolumn,showpacs,superscriptaddress]{revtex4-1}
\usepackage[american]{babel}
\usepackage[T1]{fontenc}
\usepackage[utf8]{inputenc}
\usepackage{amsmath}
\usepackage{amssymb}
\usepackage{graphicx}
\usepackage{bm}
\usepackage{color}
\usepackage[pdfborder={0 0 0}]{hyperref}
\usepackage{lineno}
\usepackage{scrpage2}
\usepackage[absolute]{textpos}
\usepackage[normalem]{ulem}

\newcommand{\vk}{\mathbf{k}}
\newcommand{\vv}{\mathbf{v}}

\newcommand{\vl}{\mathbf{\Lambda}}
\newcommand{\eps}{\varepsilon}

\newcommand{\vE}{\mathbf{E}}
\newcommand{\vB}{\mathbf{B}}
\newcommand{\vj}{\mathbf{j}}

\newcommand{\vkb}{ {\bar{\mathbf{k}}} }

\allowdisplaybreaks

\begin{document}

\title{Interrupted orbital motion in density-wave systems}

\author{Maxim Breitkreiz}
\email{breitkreiz@lorentz.leidenuniv.nl}
\affiliation{Institute of Theoretical Physics, Technische Universit\"at Dresden,
01062 Dresden, Germany}
\affiliation{Instituut-Lorentz, Universiteit Leiden, P.O. Box 9506, 2300 RA Leiden, 
The Netherlands}

\author{P. M. R. Brydon}
\affiliation{Condensed Matter Theory Center and Joint Quantum Institute,
University of Maryland, College Park, Maryland 20742, USA}
\affiliation{Department of Physics, University of Otago, PO Box
56, Dunedin 9054, New Zealand}

\author{Carsten Timm}
\email{carsten.timm@tu-dresden.de}
\affiliation{Institute of Theoretical Physics, Technische Universit\"at Dresden,
01062 Dresden, Germany}

\date{June, 17, 2016}

\begin{abstract}
In conventional metals, electronic transport in a magnetic
field is characterized by the motion of electrons along orbits on the Fermi
surface, which usually causes an increase in the resistivity
through averaging over velocities. Here we show that large
deviations from this behavior can arise in density-wave systems
close to their ordering temperature. Specifically, enhanced
scattering off collective fluctuations can lead to a
change of direction of the orbital motion on reconstructed pockets.
In weak magnetic fields, this leads to linear magnetoconductivity, 
the sign of which
depends on the electric-field direction. At a critical magnetic field, 
the conductivity crosses zero for certain directions,
signifying a thermodynamic instability of the density-wave state.
\end{abstract}

\pacs{
72.15.Lh, 
72.10.Di,
74.70.Xa,
75.30.Fv 
}

\maketitle

\section{Introduction}

The central concept in the theory of magnetotransport in metals is the
motion of electrons along the Fermi surface, driven by the
Lorentz force. This is described by the semiclassical equation of motion
\begin{equation}
\frac{d}{dt}\, \hbar\vk =  -e\, \vv_{\vk}\times \vB ,
\label{Lf}
\end{equation}
where $\hbar\vk$ is the momentum at the Fermi surface, $-e$ is the
electron charge, and $\vv_{\vk}=\nabla_{\vk}\eps_{\vk}/\hbar$ is the velocity
for the dispersion $\eps_{\vk}$.  The direction of the Lorentz force is opposite
for electron-like and hole-like Fermi pockets since $\vv_{\vk}$ is opposite. 

The driving by the Lorentz force is balanced by various scattering mechanisms, which 
also have a profound effect on the motion of the electrons in momentum space. 
Most theoretical investigations consider the case
that the scattering is approximately isotropic so that 
one can identify the electronic lifetime with the transport relaxation
time, i.e., the time needed to randomize the velocity of the electron.
In this case, the shift of the electron is obtained
by integrating Eq.\ (\ref{Lf}) over the lifetime, hence
its direction is obviously set by the Lorentz force.

The presence of {\it anisotropic} scattering significantly complicates this
picture, as we can no longer simply integrate Eq.\ (\ref{Lf}).  This
is an important problem, as strong anisotropic
scattering is expected in a number of materials of current interest, in particular in
excitonic systems such as  
tran\-si\-tion-me\-tal
dichalcogenides\cite{Rossnagel2011, Ganesh2014, Kim2015, Eom2014} and 
iron pnictides.\cite{Johnston2010, Chubukov2012,
Stewart2011, Fernandes2014, Yin2014} Here, nesting
of electron-like and hole-like Fermi pockets in the disordered
state\cite{Inosov2008, Rossnagel2011, Mazin2009}
favors the condensation of electron-hole pairs (interband excitons) due to
repulsive electronic interactions, which can drive the system into a
density-wave state with ordering vector equal to the nesting vector.\cite{Han2008,
Chubukov2008, Vorontsov2009, Brydon2009, Eremin2010, Brydon2011}

\begin{figure}[b]
\includegraphics[width=1\columnwidth]{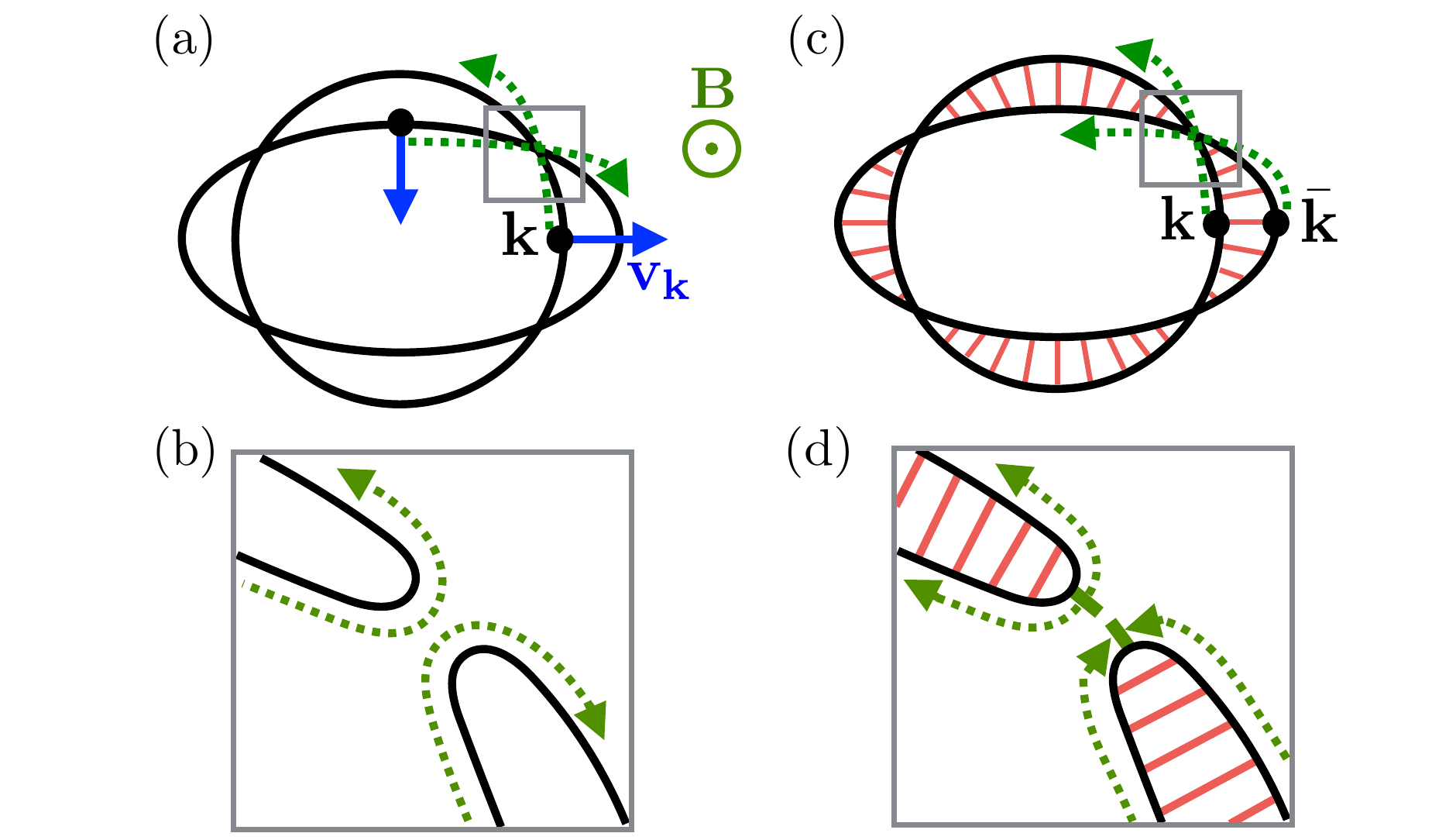}
\caption{(Color online) (a)~Nearly nested electron-like and hole-like
pockets (solid black lines) of an excitonic system in a backfolded
Brillouin zone. Dotted arrows denote the directions of orbital motion
in the absence of anisotropic scattering. (b) Close-up of the intersection, where
the reconstruction below the transition sets in. (c), (d) Same as (a), (b),
respectively, but in the presence of anisotropic scattering 
between the states $|\vk\rangle$ and $|\vkb\rangle$, indicated by
the red (gray) bars. The orbital motion on one of the pockets is
reversed. In (d), the emergence of points at which the
effective orbital motion changes direction leads to singularities.}
\label{fig:1}
\end{figure}

The Fermi pockets of a minimal model in the disordered state are sketched in
Fig.\ \ref{fig:1}(a), where the Brillouin zone has been backfolded according to 
the nesting vector. The density-wave order leads to a reconstruction of
the Fermi pockets which sets in at their intersections, as sketched in
Fig.~\ref{fig:1}(b). This results in two electron-like and two hole-like
banana-shaped pockets, which have large portions that are
dominated by the states of the original electron-like or hole-like Fermi
surfaces, and small turning regions with mixed character near the
gapped-out intersections.
Anisotropic scattering between the approximately nested parts of
the Fermi pockets is mediated by enhanced collective fluctuations,
which are particularly strong close to the transition. 
Scattering due to these fluctuations is thought to be
responsible for unconventional 
Cooper pairing\cite{Mazin2009, Chubukov2012, Graser2009, *Graser2010,
Kuroki2009, Ikeda2010, Schmiedt2014} and transport anomalies in the normal
state.\cite{Rullier-Albenque2012, Ohgushi2012,Fernandes2011, Fanfarillo2012,
Breitkreiz2013, Breitkreiz2014, Breitkreiz2014R}

In this work, we focus on the impact of anisotropic 
scattering on the orbital motion beyond the lifetime approximation. 
Due to the quasi-two-dimensional nature of many density-wave
materials, we develop our theory for a two-dimensional model with
out-of-plane magnetic field $\vB$.
Before giving an in-depth discussion, we first explain 
the results in qualitative terms. 
The strong collective fluctuations near the
density-wave transition mediate strong scattering between states
with momentum transfer close to the nesting vector.
These pairs of states, denoted by $|\vk\rangle$ and $|\vkb\rangle$, 
are connected 
by the red (gray) bars in Figs.~\ref{fig:1}(c) and \ref{fig:1}(d). The relevant
timescale for the orbital motion is set by the transport relaxation time, which
for strongly anisotropic scattering
is much longer than the timescale of the scattering
between $|\vk\rangle$ and $|\vkb\rangle$.\cite{Breitkreiz2013,Breitkreiz2014}
Thus, before integrating the equation of motion (\ref{Lf}) over the relaxation time,
the Lorentz force should first be averaged over these two states. While the 
bare Lorentz force is nearly antiparallel for $|\vk\rangle$ and $|\vkb\rangle$,
the averaged Lorentz force is the same for both. Consequently, the 
direction of the orbital motion must be reversed for one of the two 
states.

Specifically in the density-wave state, there is strong scattering
between states
on opposite sides of the same reconstructed pocket.
Consequently, a reversal of the effective orbital motion of
electrons starting on one of the sides can occur. 
This implies that the effective orbital motion changes its direction
in the turning region and so there has to exist a point where it
vanishes. This \emph{interrupted orbital motion} is our central result,
which leaves unambiguous signatures in the magnetoconductivity. 
Note that at the points of interruption, the electronic lifetimes 
are generically not
suppressed, which is a crucial difference to the previously proposed
possible interruption in systems with hot spots on the Fermi surface.\cite{Koshelev2013}

In the next section we present a detailed description of the mechanism leading to
the interrupted orbital motion.
In Sec.\ \ref{nlc} we then discuss the most dramatic
consequence of the interruption---negative longitudinal conductivity and 
the associated instability of the density-wave state. In Secs.\
\ref{two} and
\ref{four} we explicitly calculate the magnetoresistance for a two-pocket and
a four-pocket model, respectively.
We summarize our work and draw conclusions in Sec.\ \ref{sum}.

\vrule
\section{Interrupted orbital motion} \label{iom}

We now describe the interrupted orbital motion in more detail,
using the
semiclassical transport formalism.\cite{Taylor1963} Standard
approximation schemes, such as the relaxation-time approximation, are 
ill suited for the analysis of 
strongly anisotropic scattering.\cite{Pikulin2011} In this work, we
instead utilize an approximation that becomes exact in the
limit of strong anisotropy.\cite{Breitkreiz2013,Breitkreiz2014}

Our starting point is the Boltzmann equation for the
stationary non-equilibrium distribution function $f_{\vk}$,
\begin{equation}
(-e\vE - e\vv_{\vk}\times\vB)\cdot\frac{1}{\hbar} 
  \nabla_{\vk} f_{\vk}
  = \sum_{\vk'} W_{\vk}^{\vk'}\, (f_{\vk'}-f_{\vk}) ,
\end{equation}
where detailed balance requires the scattering rates
$W_{\vk}^{\vk'}$ to be symmetric in $\vk$ and $\vk'$. 
The distribution function can be
written as $f_{\vk}=n_F(\eps_{\vk})-n'_F(\eps_{\vk})\, (-e)\, \vE\cdot\vl_{\vk}$
up to linear order in the electric field
$\vE$.\cite{Price1957, *Price1958, Taylor1963} Here, $n_F(\eps_{\vk})$ is
the Fermi function and $\vl_{\vk}$ is the vector mean free path (MFP).
A standard derivation then gives\cite{Taylor1963}
\begin{equation}
\vl_{\vk} = \tau_{0,\vk} \vv_{\vk} + \tau_{0,\vk}\, \frac{e}{\hbar}
  [(\vv_{\vk}\times \vB)\cdot\nabla_{\vk}]\, \vl_{\vk}
  + \tau_{0,\vk} \sum_{\vk'} W_{\vk}^{\vk'} \vl_{\vk'} ,
\label{eq:Lk}
\end{equation}
with the lifetime
$\tau_{0,\vk} \equiv (\sum_{\vk'} W_{\vk}^{\vk'})^{-1}$.
We parametrize the momenta along a Fermi pocket by
$\vk = \vk(\alpha)$ and choose $\alpha$ to increase in the direction 
\emph{opposite} to the direction of orbital motion, which is given by
$-e\,\vv_\vk\times \vB$.
Then Eq.\ (\ref{eq:Lk}) becomes
\begin{equation}
\vl_{\vk} = \tau_{0,\vk} \vv_{\vk} + \tau_{0,\vk}\,
  \omega_{0,\vk}\, \partial_\alpha \vl_{\vk}
  + \tau_{0,\vk} \sum_{\vk'} W_{\vk}^{\vk'} \vl_{\vk'} ,
\label{eq:Lk2}
\end{equation}
where $\omega_{0,\vk} \equiv (e/\hbar)\, (\vv_{\vk}\times\vB)
\cdot \nabla_{\vk} \alpha > 0$ is the cyclotron frequency. 

For later comparison, we first consider the case of isotropic scattering.
If $W_\vk^{\vk'}$ is independent of $\vk'$ and the system satisfies
inversion symmetry, the sum in Eq.\ (\ref{eq:Lk2}) vanishes, and one obtains
\begin{equation}
\vl_{\vk}=\tau_{0,\vk}\,\big(\vv_\vk+\omega_{0,\vk}\,\partial_\alpha\vl_{\vk}
\big).
\label{eq:l}
\end{equation}
The derivative term accounts for the deviation of the MFP from its
zero-field value $\vl^{(0)}_\vk=\tau_{0,\vk} \vv_\vk$ due to the
orbital motion. This motion is
characterized by the cyclotron frequency $\omega_{0,\vk}$, which only
depends on the band parameters at $\vk$. Equation (\ref{eq:l}) is solved
by the ``Shockley tube integral''\cite{Shockley1950, Koshelev2013}
\begin{eqnarray}
\vl_{\vk} &=& \int_{\alpha}^{\infty} \!\! d\alpha' \, \vl_{\vk'}^{(0)}
  \nonumber \\
&& {}\times
  \underbrace{\frac{1}{\tau_{0,\vk'}\,\omega_{0,\vk'}}\,
   \exp\bigg(-\int_{\alpha}^{\alpha'} \!\! d\alpha''
  \frac{1}{\tau_{0,\vk''}\,\omega_{0,\vk''}}\bigg)}_{\equiv\,
  D_{0,\alpha}(\alpha')} .~
\label{eq:integ}
\end{eqnarray}
The integration over $\alpha'$ starts from
the state $|\vk\rangle$ and wraps around the pocket infinitely many
times. The expression $D_{0,\alpha}(\alpha')$ is a distribution
function over the integration range $[\alpha,\infty)$,
which corresponds to all previous states of the orbital motion ending up
at $\vk(\alpha)$. For
vanishing field (i.e., $\omega_{0,\vk}\to 0$) the weight is
concentrated at $|\vk\rangle$, while for increasing field the weight
becomes more and more evenly distributed over the pocket.

For general scattering rates, a closed-form solution of
Eq.\ (\ref{eq:Lk2}) does not exist. However, it is
possible to find an approximate solution that becomes exact in the limit
of strongly anisotropic scattering.
As the first step, we iterate Eq.\ (\ref{eq:Lk2}), which yields
\begin{equation}
\vl_{\vk} = \tau_{0,\vk} \sum_{\vk'} \pi_{\vk}(\vk')
  \left( \vv_{\vk'} + \omega_{0,\vk'}\, \partial_{\alpha'} \vl_{\vk'} \right),
\label{eq:Lk.ren}
\end{equation}
with
\begin{equation}
\pi_{\vk}(\vk') \equiv \delta_{\vk,\vk'} + W_{\vk}^{\vk'} \tau_{0,\vk'}
  + \sum_{\vk_1} W_{\vk}^{\vk_1} \tau_{0,\vk_1}
    W_{\vk_1}^{\vk'} \tau_{0,\vk'} + \ldots
\label{eq:pi.1}
\end{equation}
This is a geometric series of the matrix with entries 
$W_{\vk}^{\vk'} \tau_{0,\vk'}$. However, since this matrix has an
eigenvalue of unity, with left eigenvector
$(1,1,\ldots)$, the
series does not converge. To solve this problem, we assume inversion 
symmetry and redefine
$\pi_{\vk}(\vk')$ by subtracting the isotropic contribution to the
scattering,
\begin{eqnarray}
\lefteqn{ \pi_{\vk}(\vk') \equiv \delta_{\vk,\vk'} + W_{\vk}^{\vk'}
  \tau_{0,\vk'} - c_{\vk} } \nonumber \\
&& \quad{}+ \sum_{\vk_1} \left(W_{\vk}^{\vk_1} \tau_{0,\vk_1} - c_{\vk}\right)
    \left(W_{\vk_1}^{\vk'} \tau_{0,\vk'} - c_{\vk_1}\right)  + \ldots,\qquad
\label{eq:pi.2}
\end{eqnarray}
where $c_{\vk} \equiv \min_{\vk'}
W_{\vk}^{\vk'} \tau_{0,\vk'}$, which ensures that $\pi_{\vk}(\vk')\geq0$.
Since $\vv_{\vk}$ and $\omega_{0,\vk} \partial_\alpha
  \vl_{\vk}$ are odd under inversion, 
$c_\vk$ drops out of Eq.\ (\ref{eq:Lk.ren}).\footnote{In fact, 
inversion symmetry is not needed. Even in its absence we have 
$\sum_{\vk}\vv_{\vk}=0$ since the sum can be transformed into an
integral of $\eps_{\vk}$ over the surface of the Brillouin zone using Gau\ss{}'
theorem. This surface integral vanishes due to the periodicity of $\eps_{\vk}$.
Similarly, $\sum_{\vk} \omega_{0,\vk} \partial_\alpha \vl_{\vk}=0$ since the
expression can be decomposed into a vanishing surface integral and a term
containing $\nabla_{\vk}\times\vv_{\vk}=\nabla_{\vk}\times\nabla_{\vk}\eps_{\vk}
/\hbar=0$.}
Equation (\ref{eq:pi.2}) is again a geometric
series of matrices. Their components are non-negative and the row sums are
$\sum_{\vk}(W_{\vk}^{\vk'} \tau_{0,\vk'}-c_{\vk}) = 1 - \sum_{\vk} c_{\vk}$,
which is smaller than unity if there exists a $c_{\vk}>0$. The spectral radius
of the matrix is then smaller than unity and the series converges.
With the help of the normalization factor
$N_{\vk} \equiv \sum_{\vk'} \pi_{\vk}(\vk')$, we then define the
distribution function $P_{\vk}(\vk') \equiv \pi_{\vk}(\vk')/N_{\vk}$
and the \emph{relaxation time} $\tau_\vk\equiv N_\vk\,\tau_{0,\vk}$.
With this, Eq.\ (\ref{eq:Lk.ren}) becomes
\begin{equation}
\vl_{\vk} = \tau_{\vk} \sum_{\vk'} P_{\vk}(\vk')
  \left( \vv_{\vk'} + \omega_{0,\vk'}\, \partial_{\alpha'} \vl_{\vk'} \right),
\label{eq:Lk.ren2}
\end{equation}
which is a plausible generalization of Eq.\ (\ref{eq:l})---the
lifetime is replaced by the relaxation time and the term in
parentheses is averaged over final states.

\begin{figure}[b]
\includegraphics[width=1\columnwidth]{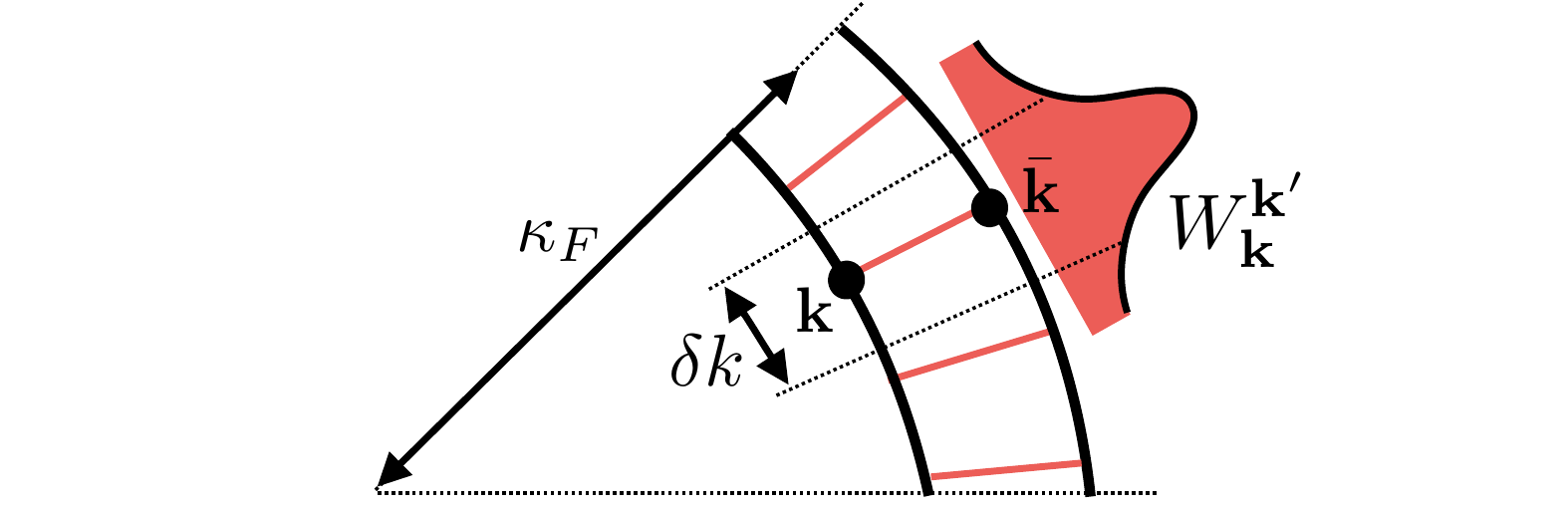}
\caption{(Color online) Sketch of a segment of a reconstructed pocket.
The scattering rate $W_{\vk}^{\vk'}$ is peaked at $\vk'=\vkb$ with
peak width $\delta k$, which is much smaller than the radius of
curvature of the pocket, $\kappa_F$.}
\label{fig:12}
\end{figure}

We now apply the general expression (\ref{eq:Lk.ren2}) to the
specific
situation in excitonic systems close to their density-wave 
instability. Here, the approximate nesting of electron and 
hole pockets enhances collective fluctuations, which mediate single-particle 
scattering between the pockets.\cite{Fernandes2011, Fanfarillo2012, Breitkreiz2014} Qualitatively, this results in a scattering rate $W_\vk^{\vk'}$ with a narrow peak at
$\vk'=\vkb$, as sketched in Fig.~\ref{fig:12}. 

Here, we consider the strongly anisotropic limit of this situation,
i.e., we assume that the peak width $\delta k$ is
much smaller than the radius of curvature of the pocket,
$\kappa_F$. We further assume that an electron repeatedly 
hopping between the states
given by the maximum of $W_\vk^{\vk'}$ is always found in one of
the two states $|\vk \rangle$ or $|\vkb \rangle$, i.e., $\bar{\vkb}=\vk$.
This is true for perfectly nested pockets, whereas imperfect nesting can
induce an additional shift along the pocket.\cite{Breitkreiz2014, Varma2001}
However, for nearly nested pockets this shift is small\cite{Breitkreiz2014}
and we therefore neglect it in the following.

For this specific form of $W_\vk^{\vk'}$, the distribution $P_\vk(\vk')$ has
narrow peaks at $\vk'=\vk$ and $\vk'=\vkb$. The leading-order
term in $\delta k/\kappa_F$ is obtained by considering $W_\vk^{\vk'}$ to be a
$\delta$-function. Including also an isotropic contribution, we write
$W_\vk^{\vk'} = w_{a,\vk} \delta_{\vk',\vkb} + w_i/N$. Here, $N$ is the
number of $\vk$ points and is included to regularize the momentum sum in the
lifetime $\tau_{0,\vk} = 1/\sum_{\vk'} W_{\vk}^{\vk'}
  = 1/(w_{a,\vk}+w_i)$. Detailed
balance requires $w_{a,\vk}=w_{a,\vkb}$. For simplicity, we assume the
strength of anisotropic scattering to be momentum independent,
$w_{a,\vk}=w_a$.\footnote{The more general case can be treated similarly.
The qualitative conclusions remain unchanged.} We then find
\begin{equation}
P_{\vk}(\vk') = \frac{w_a+w_i}{2w_a+w_i}\, \delta_{\vk',\vk}
  + \frac{w_a}{2w_a+w_i}\, \delta_{\vk',\vkb}
\end{equation}
and $\tau_{\vk} = 1/w_i$. As expected, the distribution function has two
peaks.
On the other hand, the
relaxation time is determined by the isotropic part of the scattering, which 
alone ensures a randomization of the velocity.
Equation (\ref{eq:Lk.ren2}) shows that in the highly anisotropic limit,
$w_i\ll w_a$, the MFPs $\vl_{\vk}$ and $\vl_{\vkb}$ become equal and
we obtain
\begin{equation}
\vl_{\vk} = \vl^{(0)}_\vk+\tau_{\vk}\,\omega_{\vk}\,\partial_{\alpha}\vl_{\vk},
\label{eq:lf}
\end{equation}
where the zero-field MFP reads
$\vl^{(0)}_\vk = \tau_{\vk}\, (\vv_{\vk}+\vv_{\vkb})/2$.
The effect of the magnetic field is governed by the
effective cyclotron frequency
$\omega_{\vk} \equiv (\omega_{0,\vk} - \omega_{0,\vkb})/2$, where
we employ a parametrization such that
$\partial_\alpha\bar\alpha=-1$ for convenience. Note
that since the parameter
$\alpha$ changes monotonically around the Fermi pocket, the parameter
$\bar\alpha$ belonging to $\vkb$ decreases if $\alpha$ increases.
The strong scattering between $|\vk\rangle$ and $|\vkb\rangle$ is
reflected by the mixing of the bare cyclotron frequencies $\omega_{0,\vk}$ and
$\omega_{0,\vkb}$.
The effective frequency $\omega_{\vk}$ clearly has opposite signs for
the states $|\vk\rangle$ and $|\vkb\rangle$, in contrast to the
positive bare frequencies.

Above the transition, the states $|\vk\rangle$ and $|\vkb\rangle$ lie on
two separate pockets of different (electron or hole) type,
which implies opposite signs of $\omega_{\vk}$ for
these pockets. Assuming that there are no sign changes in
$\omega_{\vk}$ within a single pocket, the solution for the MFP has the usual
form of the Shockley tube integral, Eq.\ (\ref{eq:integ}), but with
$\tau_{0,\vk}\,\omega_{0,\vk}$ replaced by $\tau_{\vk}\,\omega_{\vk}$ and
the upper limit of the integral, $\infty$, replaced by
$\infty\,\mathrm{sgn}\,\omega_{\vk}$.
For $\omega_{\vk}<0$, the direction of the integration path is thus reversed,
indicating reversed orbital motion
for electrons originating on the corresponding pocket.\cite{Breitkreiz2014}

\begin{figure}[t]
\includegraphics[width=1\columnwidth]{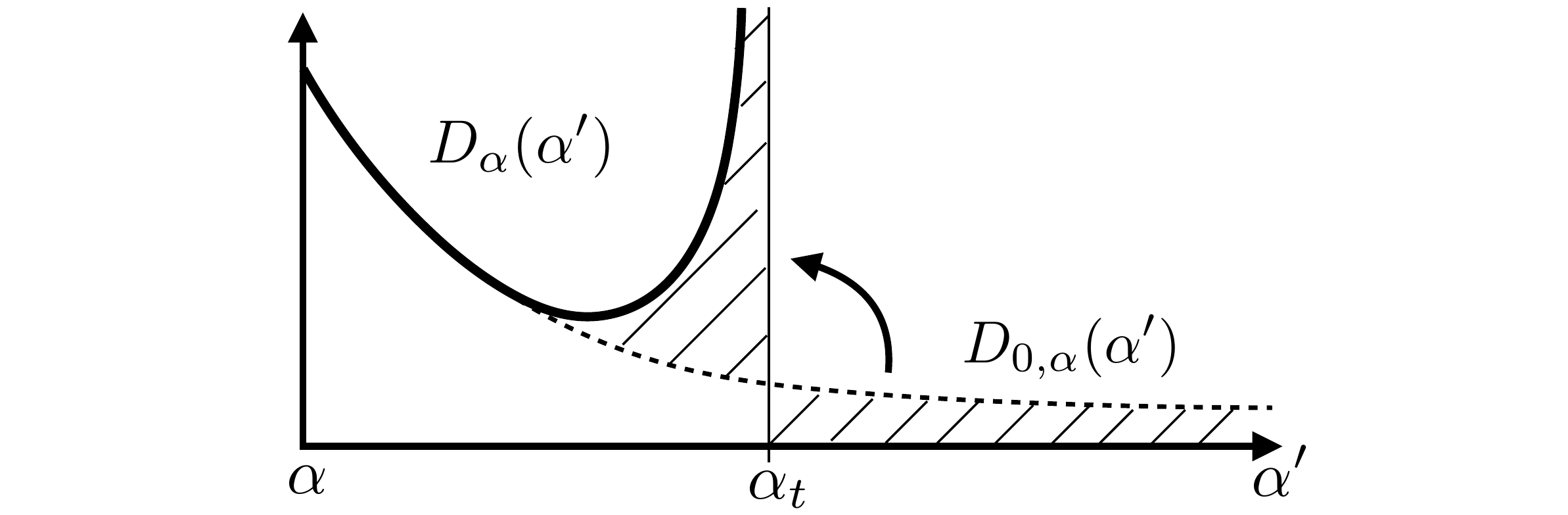}
\caption{Sketch of the distribution function $D_{\alpha}(\alpha')$
for the case of orbital motion interrupted at $\alpha_t$. The
weight, which in the absence of interruption was spread
over $[\alpha, \infty)$ ($D_{0,\alpha}(\alpha')$, dashed line), is
now restricted to the range $[\alpha,\alpha_t)$.}
\label{fig:3}
\end{figure}

The situation becomes even more interesting if there are sign changes in
$\omega_\vk$ within a single pocket. This generically happens below the
transition, as the states
$|\vk\rangle$ and $|\vkb\rangle$ now lie on the same reconstructed pocket.
At a turning point, denoted by $\vk_t$, the two peaks in $P_\vk(\vk')$
merge so that $\vk_t=\vkb_t$ and consequently $\omega_{\vk_t}=0$.
The solution of Eq.\ (\ref{eq:lf}) is then
\begin{equation}
\vl_{\vk} = \int_{\alpha}^{\alpha_t} \!\! d\alpha' \,
  \vl_{\vk'}^{(0)} \underbrace{\frac{1}{\tau_{\vk'}\,\omega_{\vk'}}\,
  \exp\bigg(-\int_{\alpha}^{\alpha'} \!\! d\alpha''
  \frac{1}{\tau_{\vk''}\,\omega_{\vk''}}\bigg)}_{\equiv\,
  D_{\alpha}(\alpha')} ,
\label{eq:integ2}
\end{equation}
where the upper limit $\alpha_t$ corresponds to the first turning point
reached from the starting point $\alpha$ backward in time
(since the integral is over previous states of an electron).
The direction of
integration is set by the sign of the effective cyclotron frequency:
$\mathrm{sgn}\,(\alpha_t-\alpha)=\mathrm{sgn}\,\omega_{\vk}$. The cutoff
$\alpha_t$ is the crucial difference to the usual tube integral, Eq.\
(\ref{eq:integ}), and signals the interrupted orbital motion. The
weight of the distribution function $D_{\alpha}(\alpha')$ is now
restricted to the range $[\alpha,\alpha_t)$,
as sketched in Fig.\ \ref{fig:3}. For strong magnetic fields,
i.e., $\tau_{\vk}\,\omega_{\vk}\gg 1$,\footnote{Note that the
quantization of cyclotron orbits plays a role if
$\tau_0 \omega > 1$. Since we consider the case
$\tau \gg \tau_0$, the limit $\tau \omega \gg 1$
is compatible with a semiclassical treatment.} the 
weight completely shifts towards the repulsive
turning point, i.e., the one the electrons move away from,
and we thus find $\vl_{\vk} \approx \vl^{(0)}_{\vk_t}$, implying that
the MFPs of all states on a single pocket approach
the MFP at the turning point. This is the 
turning point the electrons move away from.

The points of interruption are those points on the Fermi
surface where the factor $\tau_\vk\,\omega_\vk$ is zero, associated
with a sign change of $\tau_\vk\,\omega_\vk$. The crucial ingredient
for these zeros is the strongly anisotropic scattering 
between quasi-nested Fermi surfaces.
A different and, so to speak, \emph{weaker} interruption of 
orbital motion can occur in systems with hot spots on the Fermi
surface.\cite{Koshelev2013} Here, the factor $\tau_\vk\,\omega_\vk$ can
be strongly suppressed by the very short
relaxation time at hot spots, which however remains nonzero.
Since in this case $\tau_\vk\,\omega_\vk$ has no sign changes,
the tube integral has the conventional form, Eq.\ (\ref{eq:integ}),
without a cutoff.

\vrule
\section{Negative longitudinal conductivity} 
\label{nlc}

One of the most dramatic consequences of the interrupted orbital
motion is the possibility of negative longitudinal conductivity.
In the following, we discuss the general ideas before turning
to the conductivity for specific models.

The longitudinal conductivity reads
 $\sigma=\vj\cdot\vE/E^2$,
where $\vj = -2e\, N^{-1} \sum_\vk f_\vk \vv_\vk$ is the current density. To
leading order in $\vE$, we obtain
\begin{equation}
\sigma = -\frac{2e^2}{N} \sum_{\vk} n_F'(\eps_{\vk})\,
  (\vv_{\vk} \cdot \hat{\vE})\, (\vl_{\vk} \cdot \hat{\vE}) , \label{eq:sigma}
\end{equation}
with $\hat\vE\equiv \vE/E$. 
To see that the conductivity becomes negative in
certain directions above 
a critical value of the magnetic field, we consider the limit of 
 a strong magnetic field, $\tau_{\vk}\,\omega_{\vk}\gg 1$.
As discussed above, the weight of the distribution function in
Eq.\ (\ref{eq:integ2}) shifts towards the repulsive turning point,
due to the interruption of orbital motion. The 
MFPs on a single pocket become equal to $\vl^{(0)}_{\vk_t}$ and the
sum in Eq.\ (\ref{eq:sigma}) thus averages the velocities of the pocket.
For a single pocket, Eq.~(\ref{eq:sigma}) then reduces to
\begin{equation}
\sigma = 2e^2\, D(E_F)\, (\bar{\vv}\cdot\hat{\vE})\,
  \big(\vl^{(0)}_{\vk_t}\cdot\hat{\vE}\big) ,
  \label{limit}
\end{equation}
where $\bar{\vv}\equiv \sum_{\vk} n_F'(\eps_{\vk})\, \vv_{\vk}/
\sum_{\vk} n_F'(\eps_{\vk})$ and $D(E_F)$ is the density of states at the Fermi
energy. Since generically $\bar{\vv}$ and $\vl^{(0)}_{\vk_t}$ point in
different directions, this yields the surprising result that
there always exist electric-field directions 
$\hat{\vE}$ for which the conductivity is negative. While the
contributions of several Fermi pockets add up, there is no cancelation
for inversion-symmetric systems 
since both $\bar{\vv}$ and $\vl^{(0)}_{\vk_t}$ are odd under inversion.
As the conductivity is positive for vanishing magnetic field,
the strong-field limit 
Eq.~(\ref{limit}) implies a critical 
magnetic field $B_c$ at which the longitudinal conductivity changes sign
and the system becomes unstable,\cite{Landau1984}
perhaps towards a state with phase separation frustrated by the
long-range Coulomb interaction.\cite{EmK93,Tim06}
Since the latter is not included in our model, the investigation of the
new state requires further theoretical effort.

\vrule
\section{Two-pocket model} \label{two}

\begin{figure}[t]
\includegraphics[width=0.8\columnwidth]{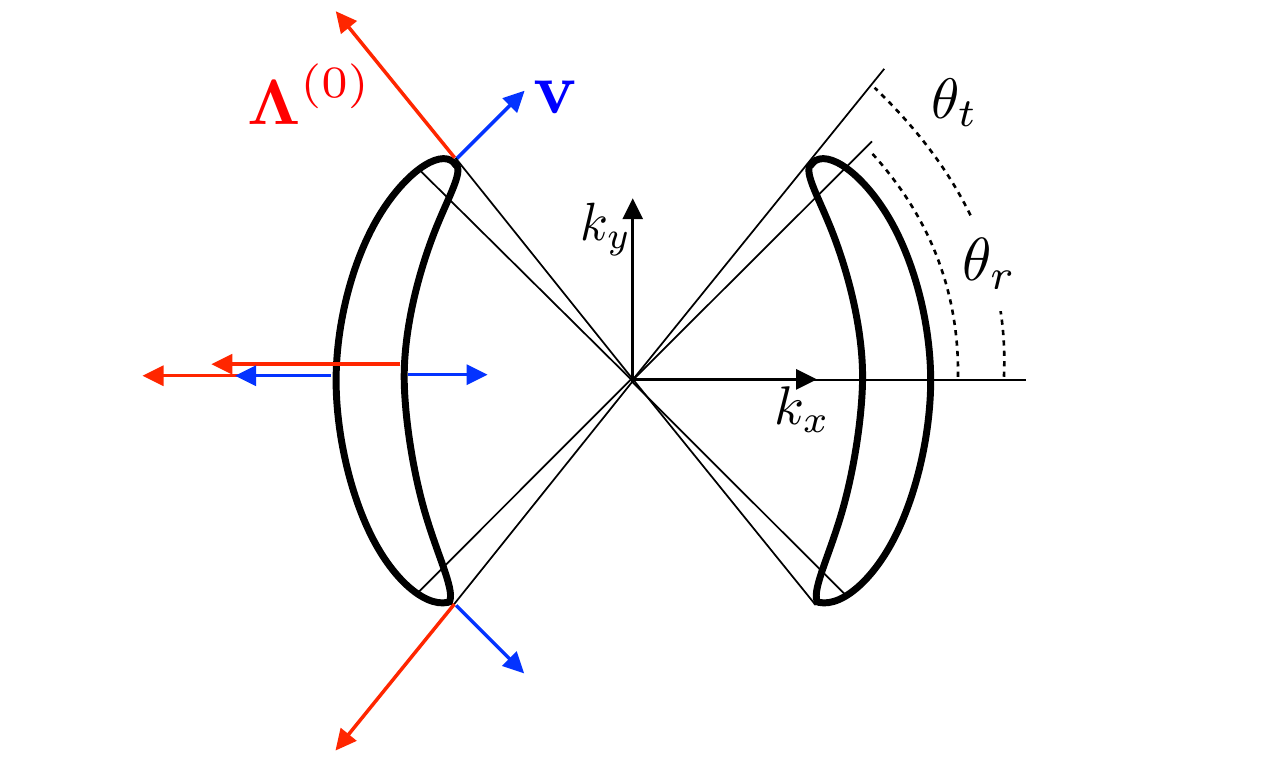}
\caption{(Color online) Fermi surface of the model used to calculate the
conductivity. Close to the transition, each reconstructed (banana-shaped)
pocket mainly consists of two segments of circles with 
different radii. For the right pocket these segments are located at the polar angles 
$-\theta_r<\theta<\theta_r$. The turning regions, found in the range
$\theta_r<\theta<\theta_t$ and equivalent ranges, are assumed to be much 
smaller than the main parts. The direction of the velocity (blue arrows)
and the zero-field MFP (red arrows) are indicated for the left pocket.}
\label{fig:s1}
\end{figure}

Our model features two equivalent reconstructed pockets, sketched in Fig.\
\ref{fig:s1}. We assume the two pockets to be electron like. The results for the
conductivity for the case of hole-like pockets are identical as the additional
sign drops out. Since each of the two pockets gives the same contribution to 
the conductivity, we focus only on the right-hand pocket in the following. We
divide the pocket into four parts:  Two large segments of circles, which
constitute the main part of the pocket,   and two turning regions, which are
assumed to be much smaller than the large segments sufficiently close to the
transition temperature. The crossover between the large parts  and the turning
regions takes place at the polar angles $\pm\theta_r$. The large segments are
found in the angular range $-\theta_r<\theta<\theta_r$ and the two turning
regions at $-\theta_t<\theta<-\theta_r$ and $\theta_r<\theta<\theta_t$. 

\vrule
 \subsection{Mean free path}

In the main part of the pocket, the velocity is parallel or antiparallel
to the momentum, depending on the side of the pocket. Assuming 
the magnitude of the velocity to be constant at each side,
we can write $\vv_\vk=v_s\,\hat{\vk}$, 
where $\hat{\vk}=\vk/|\vk|$ and $s=1,2$ denotes the side of the pocket.
$v_1$ and $v_2$ have opposite sign. The zero-field MFP
$\vl^{(0)}_\vk=\tau_\vk\,(\vv_\vk+\vv_\vkb)/2$, which is determined
by the velocity averaged over the two sides, can then
be written as $\vl^{(0)}_\vk=l\,\hat{\vk}$, where $l=(v_s+v_{\bar{s}})/(2w_i)$
and $\bar{s}=2,1$ if $s=1,2$.
 
In the turning region, the velocity direction changes by $180^\circ$
as we go along the pocket from $\theta_r$ over 
$\theta_t$ back to $\theta_r$. 
In contrast, the direction of the zero-field MFP is not reversed but only
changes by $\pm 90^\circ$ on the way from $\theta_r$ to $\theta_t$ and then
back by $\mp 90^\circ$ from $\theta_t$ to $\theta_r$.
However, this result relies on the extreme assumption of
$\delta$-function scattering. In reality, the peak in the scattering
rate $W_\vk^{\vk'}$ has a nonzero width $\delta k$, which results in
$\vl^{(0)}_\vk$ being averaged over momentum-space regions of diameter
$\delta k$. This naturally results in a smaller difference between
the directions of $\vl^{(0)}_\vk$ at $\theta_t$ and  $\theta_r$.
If the size of the turning region is small compared to $\delta k$,
the average is dominated by states from the large parts of the pockets.
In agreement with our assumption of small turning regions, we 
thus neglect the variation of $\vl^{(0)}_\vk$ in the
turning regions and
assume the validity of $\vl^{(0)}_\vk=l\,\hat{\vk}$ for the whole pocket.

We use Eq.\ (\ref{eq:integ2}) for the MFP 
$\vl_\vk$ in the presence of a
magnetic field. We now divide the integral
into a contribution from the main part and a contribution from the turning region,
\begin{widetext}
\begin{align}
\vl_{\vk} &= \int_{\alpha}^{\alpha_t} \!\! d\alpha' \,
  \vl_{\vk'}^{(0)} \frac{1}{\tau_{\vk'}\,\omega_{\vk'}}\,
  \exp\bigg({-}\int_{\alpha}^{\alpha'} \!\! d\alpha''
  \frac{1}{\tau_{\vk''}\,\omega_{\vk''}}\bigg) \\
  &= \int_{\alpha}^{\alpha_r} \!\! d\alpha' \,
  \vl_{\vk'}^{(0)} \frac{1}{\tau\,\omega}\,
  \exp\bigg({-}\frac{\alpha'-\alpha}{\tau\,\omega}\bigg) + \exp\bigg({-}\frac{\alpha_r-\alpha}{\tau\,\omega}\bigg)\,
  \int_{\alpha_r}^{\alpha_t} \!\! d\alpha' \,
  \vl_{\vk'}^{(0)} \frac{1}{\tau_{\vk'}\,\omega_{\vk'}}\,
  \exp\bigg({-}\int_{\alpha_r}^{\alpha'} \!\! d\alpha''
  \frac{1}{\tau_{\vk''}\,\omega_{\vk''}}\bigg),
\label{eq:sinteg1}
\end{align}
\end{widetext}
where we have assumed a constant product of the relaxation time and the 
effective cyclotron frequency,
$\tau_\vk\,\omega_\vk=\tau\,\omega$, in the main part of the pocket.
To leading order in $|\theta_r-\theta_t|$,
$\vl_{\vk'}^{(0)}$ can be taken out of the second integral, which then reduces to
unity, leading to the result
\begin{eqnarray}
\vl_{\vk} &=& \int_{\alpha}^{\alpha_r} \!\! d\alpha' \,
  \vl_{\vk'}^{(0)} \frac{1}{\tau\,\omega}\,
  \exp\bigg({-}\frac{\alpha'-\alpha}{\tau\,\omega}\bigg) \nonumber \\
 &&{}+ \exp\bigg({-}\frac{\alpha_r-\alpha}{\tau\,\omega}\bigg)\,
  \vl_{\vk_r}^{(0)}.
  \label{eq:sinteg2}
\end{eqnarray}
To calculate the $x$ and $y$ components of $\vl_{\vk}$ in a compact way,
we introduce a complex notation and represent a vector $\vv=(v^x,v^y)$
as $v^x+i v^y$. In this notation, the zero-field MFP,
$\vl^{(0)}_\vk=l\,\hat{\vk}$, can be written as
$\Lambda_{\vk}^{(0) x}+i \Lambda_{\vk}^{(0) y}=l\,e^{i\theta}$. 
Inserting this
into Eq.\ (\ref{eq:sinteg2}) we obtain
\begin{eqnarray}
\Lambda_{\theta}^{x}+i\Lambda_{\theta}^{y}  &=&  \frac{l}{1+(\tau\,\omega)^{2}}
\Big[e^{i\theta}(1+i\,\tau\,\omega) \nonumber \\
&&{}+\tau\,\omega\, e^{-(\theta_r-\theta)/\tau\,\omega}\,
 e^{i \theta_r} 
\big(\tau\,\omega-i\big)\Big],
 \label{eq:sLfinal}
\end{eqnarray}
where we use the polar angle $\theta$
as the parameter $\alpha$.

\vrule
\subsection{Magnetoconductivity}

\begin{figure}[t]
\includegraphics[width=1\columnwidth]{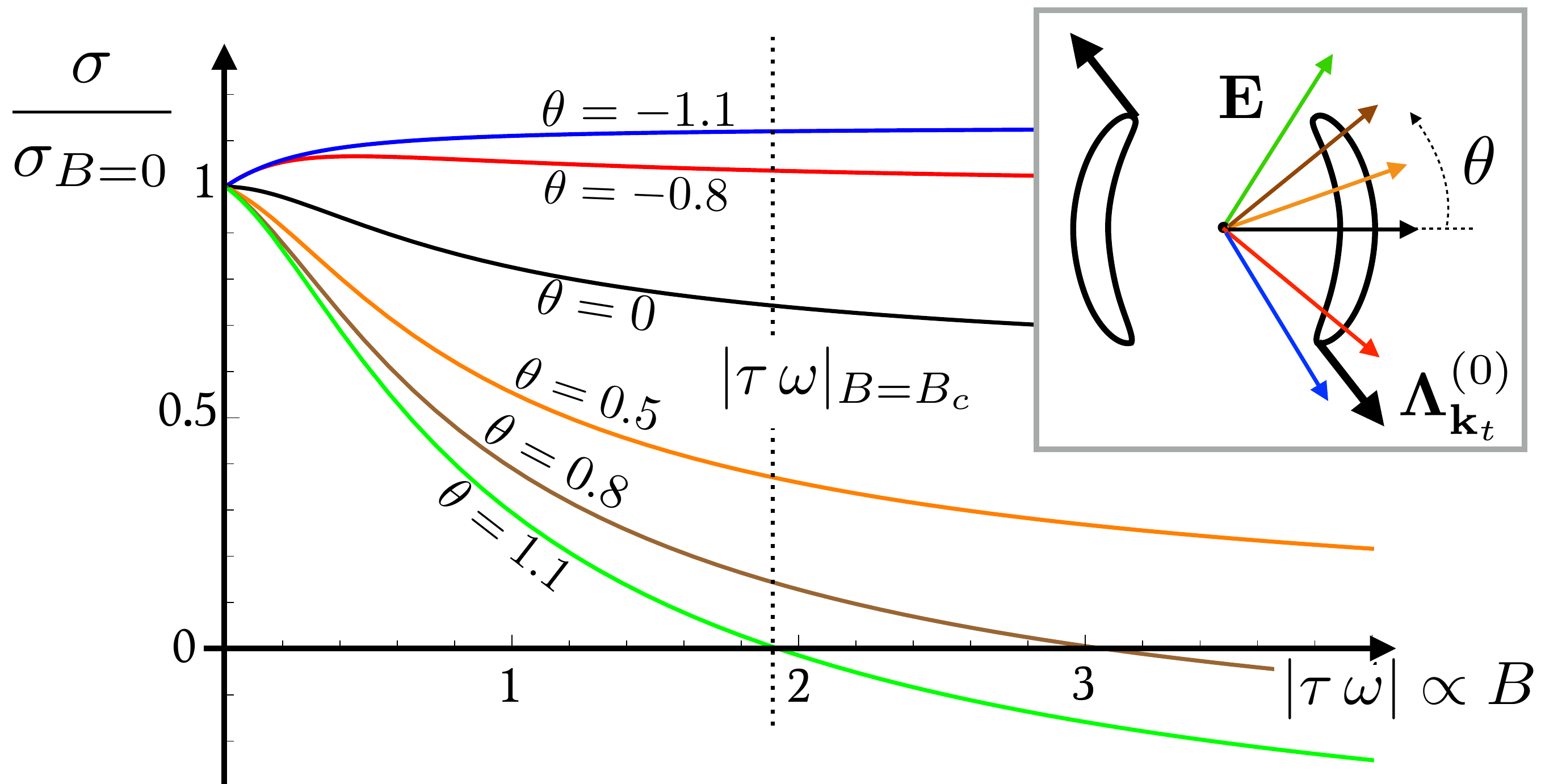}
\caption{(Color online) Magnetoconductivity for various
electric-field directions indicated in the inset. The model includes two
reconstructed pockets with turning points at $\theta=\pm\pi/3,\pi \pm
\pi/3$. $|\tau_\vk \omega_\vk|$ is set to $|\tau\,\omega|$ in the main parts
of the reconstructed pockets. The turning regions are assumed to be 
small. The zero-field MFP,
$\vl^{(0)}_{\vk}$, is taken to be parallel to the radial vector with
constant absolute value.}
\label{fig:4}
\end{figure}

The MFP in Eq.\ (\ref{eq:sLfinal}), together with the approximation
$-n_F'(\eps_\vk)\approx\delta (\eps_\vk)$, which is
valid at low temperatures, determines the conductivity given
in Eq.\ (\ref{eq:sigma}).
As shown in Fig.\ \ref{fig:4}, for small
magnetic fields the magnetoconductivity is linear and highly
anisotropic.
Most strikingly, its sign depends on the direction of the
electric field, which originates from the shift of
weight of the distribution function in Eq.~(\ref{eq:integ2}) towards
the repulsive turning points, as discussed above.

The instability occurs when the effective cyclotron frequency is on the
order of the inverse relaxation time, $1/\tau_\mathrm{k}$. For typical
metals this corresponds to a magnetic-field strength on the order  
of $1\,\mathrm{T}$, which will be significantly enhanced for
bad metals such as iron pnictides.
Additional closed Fermi surfaces, which are not reconstructed and
on which the orbital motion is not interrupted, add a positive contribution to
the total conductivity. Although the instability should
persist since the contribution from these pockets is strongly suppressed
in large fields, the critical magnetic field
will be further enhanced.

 On the other hand, for
small fields the additional
contribution  is only quadratic in 
$B$, and so the direction-dependent linear
magnetoconductivity is unaffected. This makes it the most
readily observable signature of the interrupted orbital motion.

\vrule
\section{Four-pocket model} \label{four}

Besides the two-pocket scenario, the density-wave transition might also lead to  
four reconstructed pockets---one pair of symmetry-related electron-like pockets
and one pair of symmetry-related hole-like pockets. The electron and hole pockets
are generically different in size, where
the relative size can be tuned, e.g., by doping and pressure. 
The calculation of the conductivity contribution of two additional pockets
is analogous to the previous section. For simplicity, we take the absolute
value of the velocity, the effective cyclotron frequency, and the absolute value
of the zero-field MFP to be equal for all pockets. 

\begin{figure}[t]
\includegraphics[width=\columnwidth]{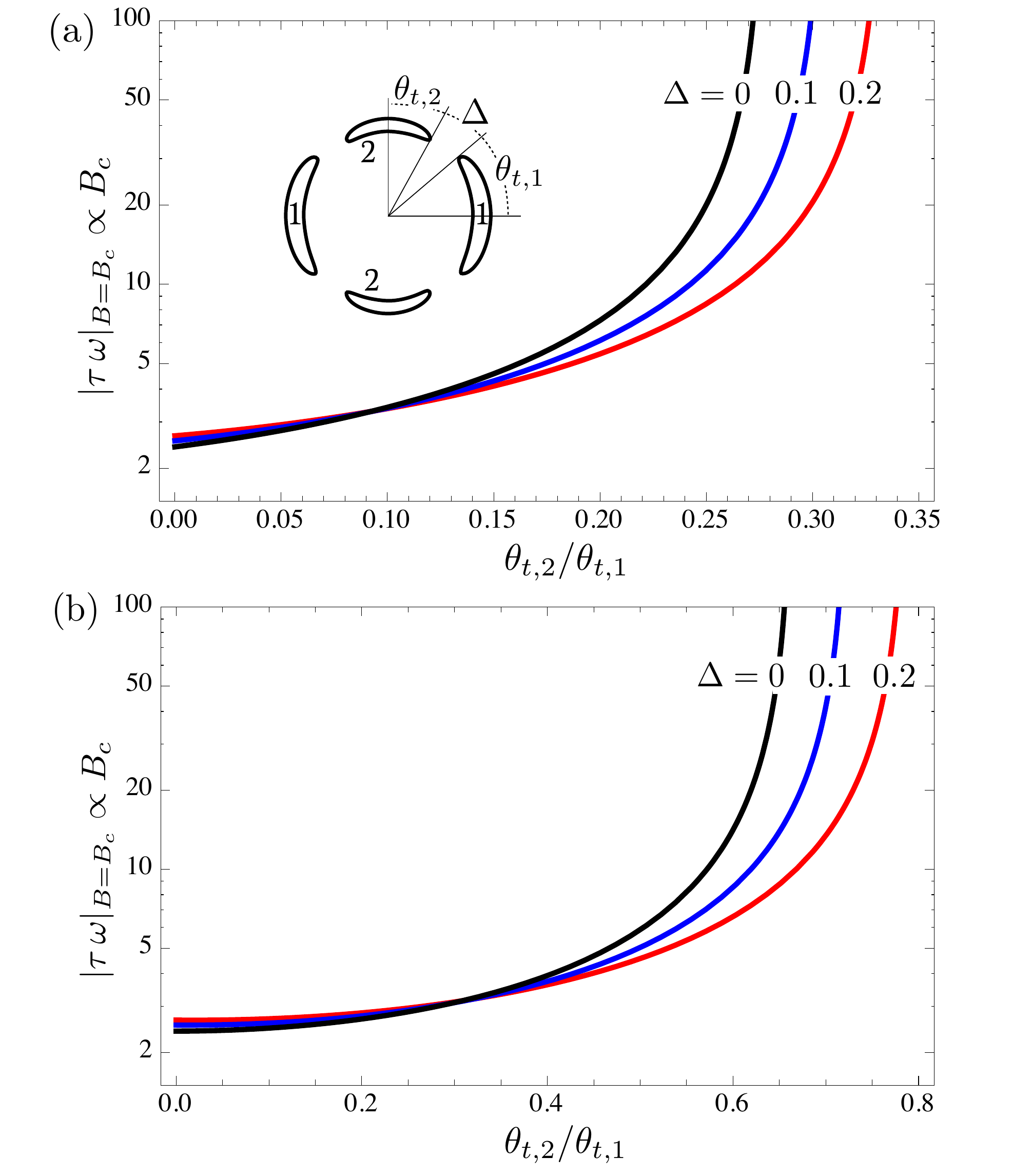}
\caption{(Color online) Effective cyclotron frequency multiplied by the relaxation time,
$|\tau\,\omega|_{B=B_c}$, as a function of the 
relative size of the pockets, $\theta_{t,2}/\theta_{t,1}$,
for various gaps $\Delta$ between the pockets.
$|\tau\,\omega|_{B=B_c}$ is a measure for the critical magnetic field,
which is proportional to this quantity.
(a) Results for equal densities of states of all pockets.
The inset shows a sketch of the Fermi surface with the relevant parameters.
(b) Results for the density of states of the first pair of pockets
(pair 1) being four times larger then that of pair 2. Note the different 
scale of the $\theta_{t,2}/\theta_{t,1}$-axis.}
\label{fig:s2}
\end{figure}

The most important parameters in the four-pocket case are the relative 
size of the two inequivalent pocket pairs and the difference between their
densities of states. As sketched in the inset of Fig.\ \ref{fig:s2},
we parametrize the two pocket sizes by the two extremal polar
angles  $\theta_{t,1}$ and $\theta_{t,2}$. The sketch also shows 
the gap between the two pockets, described by the angle
$\Delta$. We will only consider small values for $\Delta$ as we expect to find
interrupted orbital motion close to the transition temperature, where
it is small.  

The instability of the density-wave state is the most interesting consequence
of interrupted orbital motion. As shown in Fig.\ \ref{fig:4}, in the two-pocket 
model the instability occurs at magnetic fields for which the cyclotron frequency
is on the order of the inverse relaxation time, $|\tau\,\omega|_{B=B_c}\sim 1$. 
In the following, we consider how 
the critical magnetic field is modified by the two additional pockets. 
In Fig.\ \ref{fig:s2}, we plot $|\tau\,\omega|_{B=B_c}\propto B_c$
for the four-pocket case as a function of the 
relative size of the pockets, for several values of $\Delta$. The
results show, first of all, that a finite critical field $B_c$ still exists,
i.e., the instability also occurs in the case of four pockets. Compared to the 
two-pocket case, the addition of two more pockets changes $B_c$
only slightly, as long as one pair of pockets is dominant: If the pockets have the same 
density of states, one pair should be larger by approximately a factor of $5$, whereas 
if it has a four times larger density of states, it must be only twice as large.
If the pockets become close in size the critical field
increases rapidly and the instability eventually vanishes. 

\begin{figure}[t]
\includegraphics[width=\columnwidth]{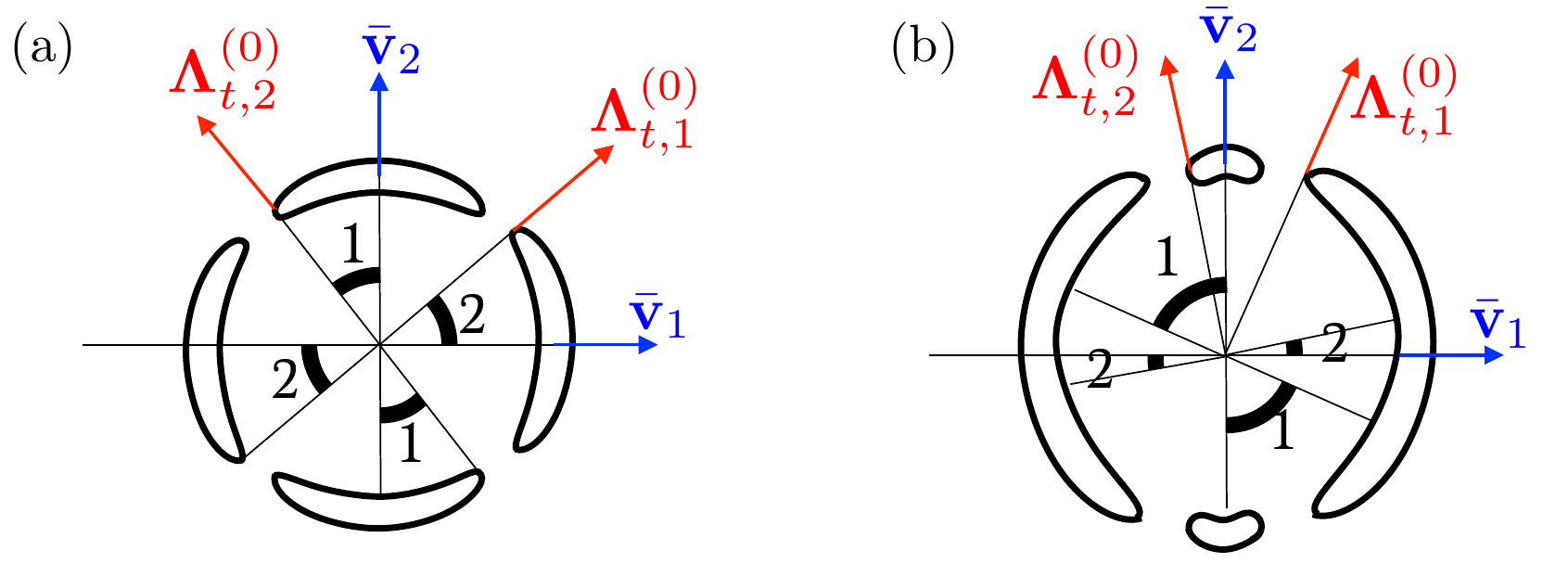}
\caption{(Color online) Angle ranges of the electric-field direction $\hat{\vE}$,
for which the first term (1)
and the second term (2) in Eq.\ (\ref{limit4}) are negative  for the case that
(a) the pairs of pockets are nearly of the same size and (b) one 
pair of pockets is significantly larger than the other.
In the second case, the two negative regions come closer to each other, which
favors negative total conductivity. }
\label{fig:s3}
\end{figure}

Qualitatively, this behavior can be understood from the 
conductivity in the limit of strong magnetic fields. The
obvious extension of Eq.\ (\ref{limit}) to the four-pocket case is
\begin{eqnarray}
\sigma &=& 2e^2\,\big[ D_1(E_F)\, (\bar{\vv}_1\cdot\hat{\vE})\,
  \big(\vl^{(0)}_{t,1}\cdot\hat{\vE}\big) \nonumber \\
  &&{}+D_2(E_F)\, (\bar{\vv}_2\cdot\hat{\vE})\,
  \big(\vl^{(0)}_{t,2}\cdot\hat{\vE}\big) \big] .
  \label{limit4}
\end{eqnarray}
According to the discussion in section \ref{nlc}, the contribution of each pair is negative 
for certain directions $\hat{\vE}$ of the electric field. These directions are indicated in 
Fig.\ \ref{fig:s3}.
Here, we assume the effective cyclotron motion to be in the \emph{same} direction,
namely counterclockwise, for the electron-like and the hole-like pockets.
This is because the effective cyclotron frequency is the difference between the
bare cyclotron frequencies for the two states $|\mathbf{k}\rangle$ and 
$|\bar{\mathbf{k}}\rangle$, which are proportional to the inverse effective masses.
It is natural to assume that either the hole or the electron band
in the \emph{disordered} state has the larger effective mass for all $\mathbf{k}$.
This means that the effective cyclotron frequency has the same sign for
\emph{both} types of reconstructed pockets. Furthermore,
the vector MFP at the turning points of both the electron-like and the
hole-like pockets is assumed to point outward. The direction of the MFP is determined
by the vector sum of velocities of the two states $|\mathbf{k}\rangle$
and $|\bar{\mathbf{k}}\rangle$, i.e., set by the larger velocity.
It is again natural to assume that either the hole or the electron band
in the disordered state has the larger Fermi velocity so that for
both types of reconstructed pockets the MFP will
point either parallel or antiparallel to the radial direction.
Whether we choose parallel or antiparallel does not matter for the conductivity 
as the sign drops out.
We observe that one of the two terms in Eq.\ (\ref{limit4}) is always positive. This 
positive term can raise the conductivity to positive values, which explains why the 
instability can be absent in the four-pocket case. However, if one pocket becomes larger
than the other
the spacing between the regions of negative contributions becomes smaller.
Then directions exist for which the contribution from the larger pockets is negative
[region 1 in Fig.\ \ref{fig:s3}(b)], while the one from the smaller pockets is positive
but smaller in magnitude since its sign change is close by, resulting in
a negative total conductivity.
This tendency is further enhanced if the larger pocket has 
a larger density of states, which increases the negative contribution.

\begin{figure}[t]
\includegraphics[width=0.9\columnwidth]{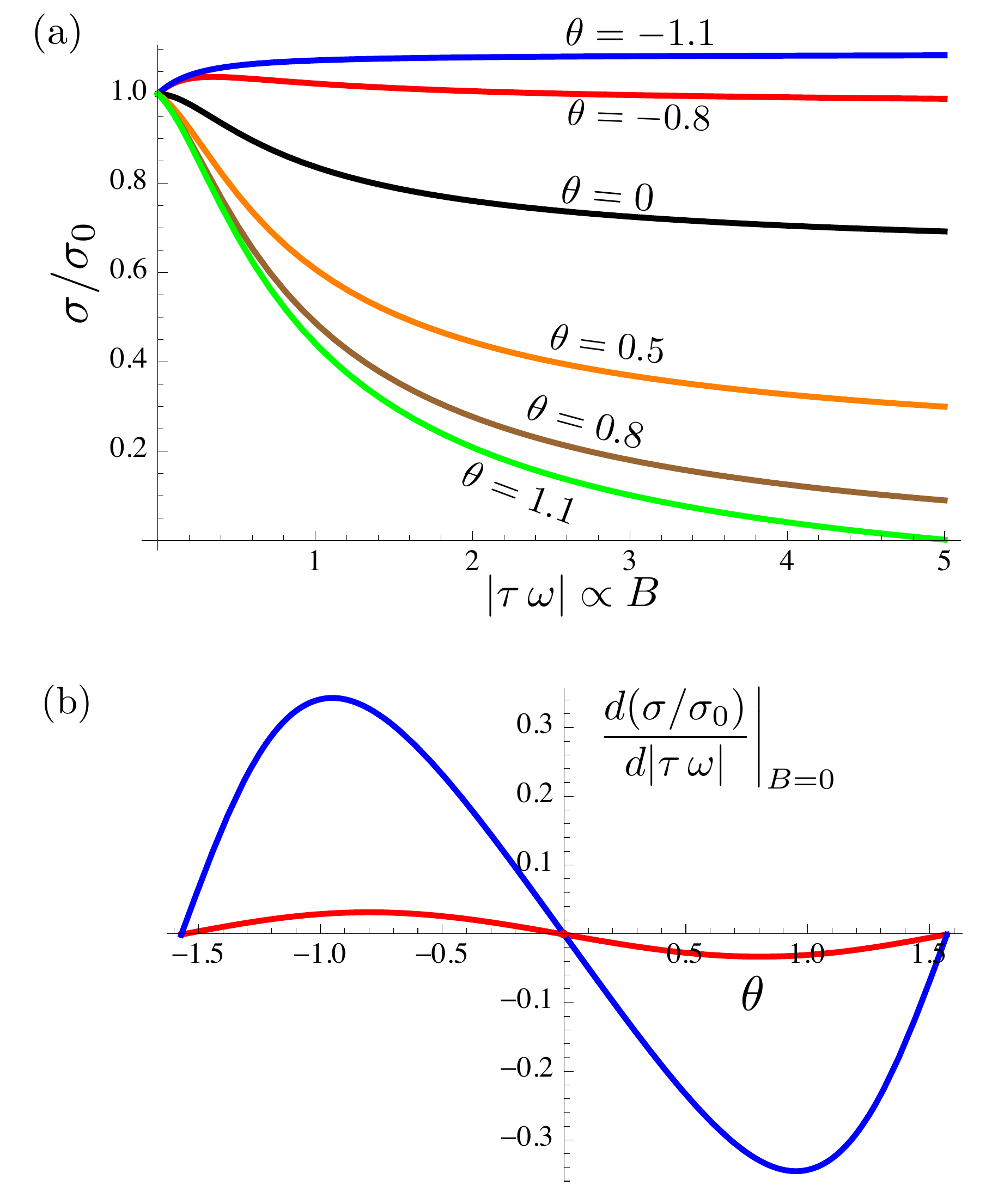}
\caption{(Color online) (a) Magnetoconductivity for various
electric-field directions, $D_1(E_F)=4\,D_2(E_F)$, 
$\Delta=0.1$, and $\theta_{t,2}/\theta_{t,1}=0.5$. (b) Differential magnetoconductivity
at $B=0$ as a function of the electric-field direction for $D_1(E_F)=4\,D_2(E_F)$
(blue curve) and  $D_1(E_F)=D_2(E_F)$ (red curve).
In both cases, $\Delta=0.1$ and $\theta_{t,2}/\theta_{t,1}=0.5$.
Note that for $D_1(E_F)=D_2(E_F)$ (red curve), one pair of pockets
dominates only weakly over the other so that no instability 
occurs, while the sign-changing linear magnetoconductivity is still present.
}
\label{fig:s4}
\end{figure}

The magnetoconductivity for the four-pocket case, shown in Fig.\ \ref{fig:s4}(a),
is qualitatively similar to the two-pocket case (cf.\ Fig.\ \ref{fig:4}). Figure \ref{fig:s4}(b) shows the differential magnetoconductivity
at $B=0$. Note that the linear contribution at low fields, 
the sign of which depends
on the electric-field direction, persists also for systems in which one
pair of pockets dominates only weakly
over the other. Although no instability occurs in this case, the 
direction dependence of the linear contribution can still serve as 
a signature of interrupted orbital motion.

\vrule
\section{Conclusions}
\label{sum}

In summary, we have investigated the magnetotransport properties in a
metal close to a density-wave instability. We find that strongly
anisotropic scattering between approximately nested Fermi pockets can
lead to reversed orbital motion of charge carriers in a magnetic field.
In the density-wave state, this generically results in points on the
Fermi surface where the orbital motion changes direction.

The interruption of the orbital motion gives rise to linear
magnetoconductivity, the sign of which depends on the
direction of the applied electric field, and an unusual instability
of the system characterized by a vanishing longitudinal conductivity
in strong magnetic fields. The latter effect, which
leads to a so-far unexplored inhomogeneous state, might be
hard to access experimentally if the system contains additional  
charge carriers with uninterrupted orbital motion, such as in iron pnictides.
More promising candidate materials are excitonic systems
without additional Fermi surfaces besides the nearly nested ones.
The effect of sign-changing linear magnetoconductivity,
on the other hand, does not require strong fields and is not 
overshadowed by the contribution of additional Fermi surfaces, which
is quadratic in the magnetic field. We expect this effect to be visible
in the direction-resolved magnetoresistance in weak magnetic fields.

\vrule
\acknowledgments

We thank J. Schmiedt,  D. Pfannkuche, D. Efremov, and A. Alt\-land for helpful
discussions. Financial support by the Deutsche Forschungsgemeinschaft through
Research Training Group GRK 1621 and Collaborative Research Center SFB 1143 is
gratefully acknowledged. P.M.R.B. acknowledges support from Microsoft Station Q,
LPS-CMTC, and JQI-NSF-PFC. This work is part of the research program of the Foundation for Fundamental Research on Matter (FOM), which is part of the Netherlands Organisation for Scientific Research (NWO).

\bibliography{refs}

\end{document}